\documentclass[sigplan,screen]{acmart}

\setcopyright{none}

\usepackage{fancyvrb}
\usepackage{fvextra}
\fvset{codes={\catcode`$=3}}
\fvset{fontsize=\smaller}
\fvset{numbersep=5pt}

\SaveVerb{evaluate}|evaluate|
\SaveVerb{diff}|diff|
\SaveVerb{reverse}|reverse|
\SaveVerb{reversec}|reversec|
\SaveVerb{Smooth}|Smooth|

\newcommand{\Vitem}{%
  \SaveVerb[aftersave={%
  \item[\UseVerb{Vitem}]}]{Vitem}}

\definecolor{mylightgray}{gray}{0.8}

\usepackage{relsize}

\usepackage{acro}
\DeclareAcronym{ad}{
  short = AD,
  long = automatic differentiation
}
\DeclareAcronym{cbpv}{
  short = CBPV,
  long = call-by-push-value
}

\usepackage{soul}
\sethlcolor{mylightgray}
\newcommand{\bhl}[1]{\hl{\mbox{\textbf{#1}}}}

\usepackage{ulem}
\usepackage{xfrac}

\title{Automatic Differentiation via Effects and Handlers}
\subtitle{An Implementation in Frank}

\author{Jesse Sigal}
\orcid{0000-0002-5117-8752}
\affiliation{%
    \institution{University of Edinburgh}
    \city{Edinburgh}
    \country{U.K.}
}
\email{jesse.sigal@ed.ac.uk}

\acmConference[PEPM'21]
    {Partial Evaluation and Program Manipulation}
    {January 18--19}
    {Online}



\begin{abstract}
\Acf{ad} is an important family of algorithms which enables derivative based optimization.
We show that \ac{ad} can be simply implemented with effects and handlers by doing so in the Frank language.
By considering how our implementation behaves in Frank's operational semantics, we show how our code performs the dynamic creation of programs during evaluation.
\end{abstract}

\begin{document}

\maketitle

\section{Introduction}
Machine learning, artificial intelligence, scientific modelling, information analysis, and other data heavy fields have driven the demand for tools which enable derivative based optimization.
The family of algorithms known as \acf{ad} is the foundation of the tools which achieve this.
The family can be coarsely divided into \textit{forward mode} and \textit{reverse mode}.
Multiple modes exist because their asymptotics depend on different features of the differentiated programs.
Forward mode \ac{ad} was introduced in 1964 by Wengert \cite{wengert_simple_1964}, and reverse mode \ac{ad} was created by Speelpenning in his 1980 thesis \cite{speelpenning_compiling_1980}.

It is not surprising that, given its history, \ac{ad} has been implemented in many different ways.
Many popular tools such as ADIFOR \cite{bischof_adifor_1996}, ADIC \cite{bischof_adic_1997}, and Tapenade \cite{pascual_tapenade_2008,hascoet_tapenade_2013} work via source transformation.
These transformations take place on languages such as C and FORTRAN, and thus all of the aforementioned tools work externally from the program being written.
We shall show here that the recent Frank language \cite{lindley_be_2017,convent_doo_2020} and its operational semantics, which leverages \textit{effects and handlers}, can be informally seen as dynamically performing partial evaluation and program manipulation.

\section{Background}

\subsection{Automatic Differentiation}
We are most interested in showing the structure of \ac{ad} algorithms, so we shall only give a short intuition for \ac{ad}.
Let $f, g \colon \mathbb{R} \to \mathbb{R}$ be smooth functions (i.e.\ infinitely differentiable at all points).
The chain rule states that $\left(f \circ g\right)'(x) = f'\left(g(x)\right) \cdot g'(x)$.
\Ac{ad} algorithms use this compositional property to incrementally calculate the derivative of an entire program one basic operation at a time during evaluation.
We refer the reader to the textbook of \citeauthor{griewank_evaluating_2008} \cite{griewank_evaluating_2008} for general knowledge and to \citeauthor{hascoet_enabling_2006} \cite{hascoet_enabling_2006} for checkpointed reverse mode, our most interesting example.

\subsection{Effects and Handlers}
Effects and handlers are a structured method of including side-effects into programs.
Algebraic effects were introduced in 2001 by \citeauthor{goos_adequacy_2001} \cite{goos_adequacy_2001} handlers for them in 2009 by \citeauthor{castagna_handlers_2009} \cite{castagna_handlers_2009}.
Effects and handlers can be viewed as an extension of the common feature of catchable exceptions.
Catching an exception terminates the program delimited by the exception handling code, but effect handlers can resume the handled code and pass a value to it.
Effects and handlers can implement many common side effects such as state, exceptions, non-determinism, logging, and input-output.

\subsection{Frank}
We will be using the Frank language to implement \ac{ad}.
Frank's typing and operational semantics are inspired by call-by-push-value \cite{levy_call-by-push-value_2003}, meaning there is a distinction between values and computations.
We note Frank has a fixed left-to-right evaluation order.
Frank combines the concepts of functions and handlers by unifying them into what Frank refers to as operators, which act by application.
However, we shall usually say handler for operators which handle effects and functions otherwise.
We shall also simplify certain aspects for ease of exposition, see \citeauthor{convent_doo_2020} \cite{convent_doo_2020} for a tutorial and details.

Let us consider a simple example of a handler for a program which uses state of type \Verb|S|.
\begin{Verbatim}
state : {S -> <State S>X -> [Console]X}
state _ x            = print "end"; x
state s <get   -> k> = print "get"; state s (k s)
state _ <put s -> k> = print "put"; state s (k unit)
\end{Verbatim}
We first explain the type of \Verb|state|.
The handler \Verb|state| takes two arguments, one of type \Verb|S| and one of type \Verb|X|.
In order for \Verb|state| to be used, the context in which it is called must support the \textit{ability} \Verb|[Console]| which is a snoc-list containing exactly one \textit{instance} \Verb|Console| of the \textit{interface} \Verb|Console| (the ability \Verb|[Console, Console]| contains two distinct instances of the same interface).
The ability \Verb|[Console]| means we can use the \textit{command} \Verb|print| defined by the interface \Verb|Console|.
In the term \Verb|state s x|, the value produced by \Verb|s| can only be computed using commands from the instances in the ability \Verb|[Console]|.
On the other hand, the value produced by \Verb|x| can use commands from \Verb|[Console, State S]|.
The value \Verb|x| has access to \Verb|State| commands because the \textit{adjustment} \Verb|<State S>| extends the ambient ability \Verb|[Console]|.
We note that the adjustment \Verb|<State S>| guarantees that \Verb|state| handles all commands of the \Verb|State| interface (\Verb|get| and \Verb|set|).
The full type of \Verb|state| includes braces, showing that \Verb|state| is a suspended computation.
Frank automatically inserts these if they are absent.

We shall briefly explain some aspects of Frank's operational semantics before we go into more detail during \ac{ad} examples.
Consider the example top-level use of \Verb|state| where semicolon is sequencing and postfix \Verb|!| is nullary function application.
\begin{Verbatim}[commandchars=\\\{\}]
2 + (state 1 (put (\emph{get!} + get!); get!))
\end{Verbatim}
The ability \Verb|[Console]| is permitted at the top-level as Frank's implementation will handle it.
As the program executes, the underlined \Verb|get| is encountered and a continuation of the program delimited by \Verb|state| is captured, namely the operator \Verb|{r -> (put (r + get!); get!)}|, and bound to \Verb|k| in the body of the second line of \Verb|state|'s definition.
Once the execution of \Verb|(put (get! + get!); get!)| finishes, the first line of \Verb|state|'s definition is matched and \Verb|state| exits.

\section{Algorithm Implementations}

We will cover the implementation of four different handlers in Frank:
\begin{description}
    \Vitem|evaluate|: the most basic handler which dispatches to builtin arithmetic operations;
    \Vitem|diff|: an implementation of forward mode \ac{ad};
    \Vitem|reverse|: an implementation of reverse mode \ac{ad} which makes use of the builtin mutable state interface; and
    \Vitem|reversec|: an implementation of checkpointed reverse mode \ac{ad} which extends \UseVerb{reverse}.
\end{description}
Each of the handlers handle the interface \UseVerb{Smooth}, which conceptually corresponds to smooth functions on the real numbers.
We only include constants, negation, addition, and multiplication for simplicity, but any number of other smooth functions could be included.
Additionally, Frank currently does not support floats, so we use integers instead, however with language support floats could be used.
\begin{Verbatim}
data Nullary = constE Int
data Unary   = negateE
data Binary  = plusE | timesE

interface Smooth X =
    ap0 : Nullary -> X
  | ap1 : Unary   -> X -> X
  | ap2 : Binary  -> X -> X -> X
\end{Verbatim}
The above definition says the \UseVerb{Smooth} interface is parameterized by \Verb|X| and has three effectful commands.
The command \Verb|apN| is the \Verb|N|-ary application of a smooth function.
Note that the nullary functions are constants.
For ease of use, we define the following helper functions.
\begin{Verbatim}
c : Int -> [Smooth X] X
c i = ap0 (constE i)

n : X -> [Smooth X] X
n x = ap1 negateE x

p : X -> X -> [Smooth X] X
p x y = ap2 plusE x y

t : X -> X -> [Smooth X] X
t x y = ap2 timesE x y
\end{Verbatim}
The operational semantics of Frank allows us to treat the above helper functions as if they were commands themselves, which we will do throughout.
We will also define helper functions for the dispatching of $n$-ary functions and their derivatives to make the similarity between different \ac{ad} modes more apparent.
\begin{Verbatim}[commandchars=\\\{\}]
op0 : Nullary ->           [Smooth X] X
op1 : Unary   -> X ->      [Smooth X] X
op2 : Binary  -> X -> X -> [Smooth X] X
op0 (constE i)     = c i
op1 negateE    x   = n x
op2 plusE      x y = p x y
op2 timesE     x y = t x y

der1  : Unary  -> X ->      [Smooth X] X
der2L : Binary -> X -> X -> [Smooth X] X
der2R : Binary -> X -> X -> [Smooth X] X
der1  negateE x   = n (c 1)   $\sfrac{d}{dx}(-x) = -1$
der2L plusE   x y = c 1       $\sfrac{d}{dx}(x + y) = 1$
der2L timesE  x y = y         $\sfrac{d}{dx}(x \cdot y) = y$
der2R plusE   x y = c 1       $\sfrac{d}{dy}(x + y) = 1$
der2R timesE  x y = x         $\sfrac{d}{dy}(x \cdot y) = x$
\end{Verbatim}

\subsection{Evaluation}

The most basic handler we will consider is the \UseVerb{evaluate} handler.
It only handles \Verb|Smooth X| where \Verb|X| is instantiated to \Verb|Int|.
\begin{Verbatim}
evaluate : <Smooth Int> X -> X
evaluate x = x
evaluate <ap0 (constE i) -> k> = evaluate (k i)
evaluate <ap1 negateE x  -> k> = evaluate (k (-x))
evaluate <ap2 plusE x y  -> k> = evaluate (k (x + y))
evaluate <ap2 timesE x y -> k> = evaluate (k (x * y))
\end{Verbatim}
In the case of \Verb|constE i|, its integer parameter \Verb|i| is returned.
Each other case of \UseVerb{evaluate} takes the integer arguments passed to the command and performs the corresponding integer operation.

The \UseVerb{evaluate} handler will always be our top-level handler, and it is the only way to remove all \Verb|Smooth| interfaces.
We shall evaluate an example program where \UseVerb{evaluate} is the top-level 
handler to illustrate how Frank executes.
We will be paying special attention to how delimited continuations are captured.
We will use underlining to show what term is currently at the focus of evaluation.

Our initial program is below, and represents the term $1 + x^3 + -y^2$ evaluated at $x = 2$ and $y = 4$, which equals $-7$ .
\begin{Verbatim}
evaluate (p (c 1) (p (t (t 2 2) 2) (n (t 4 4))))
\end{Verbatim}
The current focus of evaluation is the command \Verb|c 1|.
\begin{Verbatim}[commandchars=\\\{\}]
evaluate (p \emph{(c 1)} (p (t (t 2 2) 2) (n (t 4 4))))
\end{Verbatim}
The argument \Verb|1| is in normal form (fully evaluated).
Therefore, we can handle the command \Verb|c 1|.
The handling process begins by capturing the proper delimited continuation by incrementally freezing the stack of evaluation frames.
We represent freezing by highlighting and boldface.
\begin{Verbatim}[commandchars=\\\{\}]
evaluate (p \bhl{\emph{(c 1)}} (p (t (t 2 2) 2) (n (t 4 4))))
\end{Verbatim}
\begin{Verbatim}[commandchars=\\\{\}]
evaluate \bhl{(p \emph{(c 1)} (p (t (t 2 2) 2) (n (t 4 4))))}
\end{Verbatim}
We have now reached a handler, \UseVerb{evaluate}, for the command in focus.
The frozen command (highlighted) is the captured delimited continuation.
The \Verb|ap0| case of \UseVerb{evaluate} is then matched to the command \Verb|c 1|, where \Verb|k| is bound to the continuation with \Verb|c 1| removed and \Verb|i| is bound to \Verb|1|.
The bound variables \Verb|k| and \Verb|i| are then substituted into the corresponding body of \UseVerb{evaluate}. 
\begin{Verbatim}[commandchars=\\\{\}]
evaluate \emph{(\{x -> (p x (p (t (t 2 2) 2) (n (t 4 4))))\} 1)}
\end{Verbatim}
The next step applies the continuation to \Verb|1|.
\begin{Verbatim}
evaluate (p 1 (p (t (t 2 2) 2) (n (t 4 4))))
\end{Verbatim}
The focus of evaluation now moves to \Verb|t 2 2|, and a new delimited continuation is dynamically captured.
\begin{Verbatim}[commandchars=\\\{\}]
evaluate (p 1 (p (t \emph{(t 2 2)} 2) (n (t 4 4))))
\end{Verbatim}
\begin{Verbatim}[commandchars=\\\{\}]
evaluate (p 1 (p (t \bhl{\emph{(t 2 2)}} 2) (n (t 4 4))))
\end{Verbatim}
\begin{Verbatim}[commandchars=\\\{\}]
evaluate (p 1 (p \bhl{(t \emph{(t 2 2)} 2)} (n (t 4 4))))
\end{Verbatim}
\begin{Verbatim}[commandchars=\\\{\}]
evaluate (p 1 \bhl{(p (t \emph{(t 2 2)} 2) (n (t 4 4)))})
\end{Verbatim}
\begin{Verbatim}[commandchars=\\\{\}]
evaluate \bhl{(p 1 (p (t \emph{(t 2 2)} 2) (n (t 4 4))))}
\end{Verbatim}
We have now again reached the \UseVerb{evaluate} handler, and this time match the \Verb|ap2| case, resulting in the following.
\begin{Verbatim}[commandchars=\\\{\}]
evaluate (\{x -> (p 1 (p (t x 2) (n (t 4 4))))\} \emph{(2 * 2)})
\end{Verbatim}
\begin{Verbatim}[commandchars=\\\{\}]
evaluate \emph{(\{x -> (p 1 (p (t x 2) (n (t 4 4))))\} 4)}
\end{Verbatim}
\begin{Verbatim}[commandchars=\\\{\}]
evaluate (p 1 (p (t 4 2) (n (t 4 4))))
\end{Verbatim}
Evaluation will continue as such until the final answer of $-7$ is calculated.

We have now seen how the \UseVerb{evaluate} handler interprets \UseVerb{Smooth} commands with the builtin arithmetic operations.
Even though \UseVerb{evaluate} is simple, it allows us to write our other handlers in a polymorphic fashion independent of \Verb|Int|.

\subsection{Forward mode}

Our next handler is the \UseVerb{diff} handler, which implements forward mode \ac{ad} via a method known as \textit{dual numbers}.
A dual number is a pair of real numbers where the second number represents the derivative of the first.
The \UseVerb{diff} handler handles commands with dual number arguments.
The mathematical justification of \ac{ad} is not our focus, and thus we shall just focus on the patterns of computation present without proving their correctness.
We define the \Verb|Dual| datatype and \UseVerb{diff} below.
\begin{Verbatim}
data Dual X = dual X X

v : Dual X -> X
v (dual x _) = x

dv : Dual X -> X
dv (dual _ dx) = dx

diff : <Smooth (Dual X)> Y -> [Smooth X] Y
diff x = x
diff <ap0 n -> k> =
  let r = dual (op0 n) (c 0) in
  diff (k r)
diff <ap1 u (dual x dx) -> k> =
  let r = dual (op1 u x) (t (der1 u x) dx) in
  diff (k r)
diff <ap2 b (dual x dx) (dual y dy) -> k> =
  let r = dual (op2 b x y) (p (t (der2L b x y) dx)
                              (t (der2R b x y) dy)) in
  diff (k r)
\end{Verbatim}
Notice the similarities between each of the \Verb|apN| cases.
The command being handled by \UseVerb{diff} is evaluated with \Verb|opN| in the first component of \Verb|Dual|, and a calculation involving derivatives creates the second component.

We will evaluate an example program similar to our previous one.
The program will represent the same mathematical term $1 + x^3 + -y^2$ evaluated at $x = 2$ and $y = 4$, but additionally we shall be calculating the derivative with respect to $x$ at this point, which is $12$.
This is achieved by setting $x$ to \Verb|dual 2 1| and $y$ to \Verb|dual 4 0|, where $x$ has its second component set to $1$ to treat it as the differentiated variable and $y$ has its second component set to $0$ to treat it as a constant.
\begin{Verbatim}
evaluate (diff (
  p (c 1) (p (t (t (dual 2 1) (dual 2 1)) (dual 2 1))
             (n (t (dual 4 0) (dual 4 0))))
))
\end{Verbatim}

Evaluation begins as before, with the \Verb|c 1| command being in focus and a delimited continuation being captured.
\begin{Verbatim}[commandchars=\\\{\}]
evaluate (diff (
  \bhl{p \emph{(c 1)} (p (t (t (dual 2 1) (dual 2 1)) (dual 2 1))}
  \bhl{           (n (t (dual 4 0) (dual 4 0))))}
))
\end{Verbatim}
Note how the continuation captured is delimited by \UseVerb{diff} and not \UseVerb{evaluate}.
This behavior is due to the effect typing system of Frank.
There are two different instances of the \UseVerb{Smooth} interface available to the portion of the program being handled.
By default, the innermost handler provides the instance being used by extending the ambient ability with an adaptor.
As we shall see later, Frank provides constructs allowing us to select handlers other than the innermost.
The top case of \UseVerb{diff} is matched by \Verb|c 1| with the following result.
\begin{Verbatim}[commandchars=\\\{\}]
evaluate (
  let r = dual \emph{(op0 (constE 1))} (c 0) in
  diff (
    \{x -> (p x (p (t (t (dual 2 1) (dual 2 1)) (dual 2 1))
                  (n (t (dual 4 0) (dual 4 0)))))\} r)
)
\end{Verbatim}
\begin{Verbatim}[commandchars=\\\{\}]
evaluate (
  let r = dual \uline{(c 1)} \dashuline{(c 0)} in
  diff (
    \{x -> (p x (p (t (t (dual 2 1) (dual 2 1)) (dual 2 1))
                  (n (t (dual 4 0) (dual 4 0)))))\} r)
)
\end{Verbatim}
We now have two \Verb|c| commands which will be be handled by \UseVerb{evaluate}, producing \Verb|dual 1 0| for \Verb|r|'s value.
After handling, \Verb|r| will be be substituted and the continuation applied.
\begin{Verbatim}[commandchars=\\\{\}]
evaluate (diff (
  p (dual 1 0) (p (t \emph{(t (dual 2 1) (dual 2 1))} (dual 2 1))
                  (n (t (dual 4 0) (dual 4 0))))
))
\end{Verbatim}
Evaluation will then continue in a similar manner for all remaining commands.
Each command will first be handled by \UseVerb{diff}, and the commands in the body of each \UseVerb{diff} case handled by \UseVerb{evaluate}, eventually producing \Verb|dual -7 12|.

We will now focus on Frank's ability to dynamically determine which handler handles a command.
First, we define two auxiliary functions.
\begin{Verbatim}
lift : X -> [Smooth X, Smooth (Dual X)] (Dual X)
lift x = dual x (<Smooth> (c 0))

d : {(Dual X) -> [Smooth X, Smooth (Dual X)] (Dual X)}
 -> X -> [Smooth X] X
d f x = dv (diff (f (dual x (<Smooth> (c 1)))))
\end{Verbatim}
The \textit{adaptor} \Verb|<Smooth>| in \Verb|lift| causes the command \Verb|c 0| to be associated with \Verb|Smooth X| and not the rightmost instance \Verb|Smooth (Dual X)|.
The \Verb|d| function returns the derivative of a unary function and \Verb|lift| will enable us to nest \Verb|d|.
Note that as in \Verb|lift|, \Verb|<Smooth>| in \Verb|d| causes the command \Verb|c 1| to be associated with \Verb|Smooth X|.
Consider the expression $\sfrac{d}{dx}(x \cdot \sfrac{d}{dy}(x + y)|_{y = 1})|_{x = 1}$ (which equals $1$).
The corresponding program requires \Verb|lift|.
\begin{Verbatim}[commandchars=\\\{\}]
evaluate (d \{x -> t x (d \{y -> p (lift x) y\} (c 1))\} \emph{(c 1)})
\end{Verbatim}
We evaluate until the delimited continuation is captured.
\begin{Verbatim}[commandchars=\\\{\}]
evaluate \emph{(d \{x -> t x (d \{y -> p (lift x) y\} (c 1))\} 1)}
\end{Verbatim}
\begin{Verbatim}[commandchars=\\\{\}]
evaluate (dv (diff (
  \{x -> t x (d \{y -> p (lift x) y\} (c 1))\}
  (dual 1 (<Smooth> \emph{(c 1)}))
)))
\end{Verbatim}
\begin{Verbatim}[commandchars=\\\{\}]
evaluate \bhl{(dv (diff (}
\bhl{  \{x -\textgreater t x (d \{y -\textgreater p (lift x) y\} (c 1))\}}
\bhl{  (dual 1 (\textless{}Smooth\textgreater \emph{(c 1)}))}
\bhl{)))}
\end{Verbatim}
The continuation for \Verb|c 1| is delimited by \UseVerb{evaluate} due to \Verb|<Smooth>|.
We conclude by noting that Frank will reject the nested program if \Verb|lift| is not present.

\subsection{Reverse mode}

The \UseVerb{evaluate} and \UseVerb{diff} handlers manipulate programs by capturing delimited continuations, but only in quite simple ways.
They each eventually compute a value based on the command being handled and then continue with the original program with the computed value substituted in.
The \UseVerb{reverse} handler will be different, and will build up a secondary program during the evaluation of the initial program.

Reverse mode \ac{ad} works by creating a mutable cell for each value which accumulates contributions to its derivative.
The method of accumulation is a generalized version of the \textit{backpropagation} algorithm made prominent by machine learning.
We define the datatype \Verb|Prop| for backpropagation where \Verb|Ref X| is a reference to a mutable cell containing a value of type \Verb|X|.
The \UseVerb{reverse} handler handles commands containing \Verb|Prop|'s.
\begin{Verbatim}
data Prop X = prop X (Ref X)

fwd : Prop X -> X
fwd (prop x _) = x

deriv : Prop X -> Ref X
deriv (prop _ r) = r

reverse : <Smooth (Prop X)> Unit -> [RefState, Smooth X] Unit
reverse x = x
reverse <ap0 n -> k> =
  let r = prop (op0 n) (new (c 0)) in
  reverse (k r)
reverse <ap1 u (prop x dx) -> k> =
  let r = prop (op1 u x) (new (c 0)) in
  reverse (k r);
  write dx (p (read dx) (t (der1 u x) (read (deriv r))))
reverse <ap2 b (prop x dx) (prop y dy) -> k> =
  let r = prop (op2 b x y) (new (c 0)) in
  reverse (k r);
  write dx (p (read dx) (t (der2L b x y) (read (deriv r))));
  write dy (p (read dy) (t (der2R b x y) (read (deriv r))))
\end{Verbatim}
The \UseVerb{reverse} handler makes use of the same \Verb|op| and \Verb|der| functions as \UseVerb{diff}, but is different from \UseVerb{evaluate} and \UseVerb{diff} in two important ways.
Firstly, the type of \UseVerb{reverse} shows that it requires access to the \Verb|RefState| interface of mutable state (a builtin effect of Frank that can be handled by the language implementation).
Secondly, the body of the \Verb|ap1| and \Verb|ap2| cases contains code after the use of the captured delimited continuation \Verb|k|.
We shall see these writes to memory will form the secondary program which actually accumulates derivatives.

To properly calculate derivatives with \UseVerb{reverse}, we require a small helper function which starts the process of backpropagation, which we call \Verb|grad| for gradient.
\begin{Verbatim}
grad : {(Prop X)
      -> [RefState, Smooth X, Smooth (Prop X)] (Prop X)}
    -> X -> [RefState, Smooth X] X
grad f x =
  let z = prop x (new (c 0)) in
  reverse (write (deriv (f z)) (<Smooth> (c 1)));
  read (deriv z)
\end{Verbatim}
We evaluate the same term as before.
\begin{Verbatim}[commandchars=\\\{\}]
evaluate \uline{(grad (\{x ->}
\uline{  let y = c 4 in p (c 1) (p (t (t x x) x) (n (t y y)))}
\uline{\} 2)})
\end{Verbatim}
\begin{Verbatim}[commandchars=\\\{\}]
evaluate (
  let z = prop 2 \dashuline{(new \uline{(c 0)})} in
  reverse (write (deriv (\{x ->
    let y = c 4 in p (c 1) (p (t (t x x) x) (n (t y y)))
  \} z)) (<Smooth> (c 1)));
  read (deriv z))
\end{Verbatim}
The term \Verb|new (c 0)| is handled first by \UseVerb{evaluate} for \Verb|c 0| (returning \Verb|0|), and the command \Verb|new 0| is handled by the Frank implementation and returns a new reference \Verb|<z>| whose cell contains \Verb|0|.
The result is then substituted for \Verb|z|.
\begin{Verbatim}[commandchars=\\\{\}]
evaluate (
  reverse (write (deriv \emph{(\{x ->}
  \emph{  let y = c 4 in p (c 1) (p (t (t x x) x) (n (t y y)))}
  \emph{\} (prop 2 <z>))}) (<Smooth> (c 1)));
  read (deriv (prop 2 <z>)))
\end{Verbatim}
Next, the anonymous function is applied to \Verb|prop 2 <z>|.
\begin{Verbatim}[commandchars=\\\{\}]
evaluate (
  reverse (write (deriv (
    let y = \emph{c 4} in
    p \dashuline{(c 1)} (p (t (t (prop 2 <z>) (prop 2 <z>)) (prop 2 <z>))
               (n (t y y)))
  )) (<Smooth> (c 1)));
  read (deriv (prop 2 <z>)))
\end{Verbatim}
The command \Verb|c 4| is handled by the \Verb|ap0| case of \UseVerb{reverse}, which as before creates a new reference \Verb|<r1>|, and thus \Verb|y| is substituted by \Verb|prop 4 <r1>|.
The command \Verb|c 1| will create \Verb|<r2>|.
\begin{Verbatim}[commandchars=\\\{\}]
evaluate (
  reverse (write (deriv (
    p (prop 1 <r2>)
      (p (t \emph{(t (prop 2 <z>) (prop 2 <z>))} (prop 2 <z>))
         (n (t (prop 4 <r1>) (prop 4 <r1>))))
  )) (<Smooth> (c 1)));
  read (deriv (prop 2 <z>)))
\end{Verbatim}
We have now reached the first interesting command, which matches the \Verb|ap2| case of \UseVerb{reverse}.
The captured delimited continuation is now explicitly highlighted.
\begin{Verbatim}[commandchars=\\\{\}]
evaluate (
  reverse \bhl{(write (deriv (}
  \bhl{  p (prop 1 \textless{}r2\textgreater{})                                    }
  \bhl{    (p (t \emph{(t (prop 2 \textless{}z\textgreater{}) (prop 2 \textless{}z\textgreater{}))} (prop 2 \textless{}z\textgreater{}))}
  \bhl{       (n (t (prop 4 \textless{}r1\textgreater{}) (prop 4 \textless{}r1\textgreater{}))))         }
  \bhl{)) (\textless{}Smooth\textgreater{} (c 1)))};
  read (deriv (prop 2 <z>)))
\end{Verbatim}
The result of \UseVerb{reverse} handling the command produces a new reference \Verb|<r3>|.
\begin{Verbatim}
evaluate (
  reverse (write (deriv (
    p (prop 1 <r2>)
      (p (t (prop 4 <r3>) (prop 2 <z>))
         (n (t (prop 4 <r1>) (prop 4 <r1>))))
  )) (<Smooth> (c 1)));
  write <z> (p (read <z>)
    (t (der2L timesE 2 2) (read (deriv (prop 4 <r3>)))));
  write <z> (p (read <z>)
    (t (der2R timesE 2 2) (read (deriv (prop 4 <r3>)))));
  read (deriv (prop 2 <z>)))
\end{Verbatim}
We see that the evaluation of the initial program has produced new expressions to be evaluated after the initial program finishes.
The handling by \UseVerb{reverse} will eventually handle all commands meant for it, producing the following.
\begin{Verbatim}
evaluate (
  reverse (write <r8> (<Smooth> (c 1)));
  write <r2> (p (read <r2>) 
    (t (der2L plusE 1 -8) (read (deriv (prop -7 <r8>)))));
  write <r7> (p (read <r7>) 
    (t (der2R plusE 1 -8) (read (deriv (prop -7 <r8>)))));
  write <r4> (p (read <r4>) 
    (t (der2L plusE 8 -16) (read (deriv (prop -8 <r7>)))));
  write <r6> (p (read <r6>) 
    (t (der2R plusE 8 -16) (read (deriv (prop -8 <r7>)))));
  write <r5> (p (read <r5>) 
    (t (der1 negateE 16) (read (deriv (prop -16 <r6>)))));
  write <r1> (p (read <r1>) 
    (t (der2L timesE 4 4) (read (deriv (prop 16 <r5>)))));
  write <r1> (p (read <r1>) 
    (t (der2R timesE 4 4) (read (deriv (prop 16 <r5>)))));
  write <r3> (p (read <r3>) 
    (t (der2L timesE 4 2) (read (deriv (prop 8 <r4>)))));
  write <z> (p (read <z>) 
    (t (der2R timesE 4 2) (read (deriv (prop 8 <r4>)))));
  write <z> (p (read <z>) 
    (t (der2L timesE 2 2) (read (deriv (prop 4 <r3>)))));
  write <z> (p (read <z>) 
    (t (der2R timesE 2 2) (read (deriv (prop 4 <r3>)))));
  read (deriv (prop 2 <z>)))
\end{Verbatim}
The above code is the secondary program created by \UseVerb{reverse}, which performs backpropagation.
Furthermore, if a user wished to capture this secondary program, the definition of \UseVerb{reverse} could be changed to return a suspended computation.
Thus, we could also partially evaluate the whole program (initial and backpropagation) by running only the initial program and capturing the backpropagation computation.

It could also be possible to use multi-stage programming by reifying the initial and secondary programs as a computation graph in the style of \citeauthor{wang_demystifying_2019} \cite{wang_demystifying_2019}.
Their approach uses delimited continuations and combines normal execution with building an intermediate representation.
As effects and handlers are essentially a structured use of delimited continuations, a similar story for Frank may be possible.

\subsection{Checkpointed reverse mode}

The final algorithm we shall cover is checkpointed reverse mode.
Reverse mode has maximum memory residency proportional to the number of operations (as seen in the definition of \UseVerb{reverse}).
Checkpointed reverse mode allows a trade-off between space and time by recomputing checkpointed subprograms, once without allocating memory and an additional time with memory.
However, any memory allocated in between these two runs can be safely deallocated, as it corresponds to code after the checkpointed subprogram in the original program, thus reducing maximum memory residency.

To define our new handler, we introduce a \Verb|Checkpoint| effect which takes a suspended computation that will be run multiple times.
We also define a simple \UseVerb{evaluate} like handler, \Verb|evaluatet| (see appendix for definition).
\begin{Verbatim}
interface Checkpoint X =
  checkpoint :
    {[Checkpoint X , Smooth (Prop X)] Prop X} -> Prop X
\end{Verbatim}
Frank also contains a catch-all pattern match \Verb|<m>| which matches values and commands not handled above it.
We use this feature to extend \UseVerb{reverse} by delegating any \UseVerb{Smooth} commands received to \UseVerb{reverse} and only adding a case for \Verb|checkpoint|.
\begin{Verbatim}
reversec : <Checkpoint X, Smooth (Prop X)> Unit
        -> [RefState, Smooth X] Unit
reversec x = x
reversec <checkpoint p -> k> =
  let s = new (c 0) in
  let res = <RefState> (evaluatet s (
    <Smooth(s a b -> s b)> p!
  )) in
  let r = prop (fwd res) (new (c 0)) in
  reversec (k r);
  reversec (write 
    (deriv (<Smooth(s a b -> s b), RefState> p!)) 
    (read (deriv r)))
reversec <m> = reversec (<Smooth(s a -> s)> (reverse m!))
\end{Verbatim}
Note how the checkpointed subprogram (the suspended computation \Verb|p| which is the argument of \Verb|checkpoint|) is called twice, once with \Verb|evaluatet| as the handler and once with \Verb|reversec| as the handler.
Additionally, the last case will match every \UseVerb{Smooth} command, and then reinvoke the captured computation with a new \UseVerb{reverse} handler to handle the command.

Consider the following program where \Verb|gradc| is \Verb|grad| with \UseVerb{reversec} in the place of \UseVerb{reverse}.
\begin{Verbatim}[commandchars=\\\{\}]
evaluate (gradc (\{x ->
  let y = c 2 in
  let z = \emph{checkpoint \{p x y\}} in
  let a = checkpoint \{let w = checkpoint \{t x z\} in p w y\} in
  p a x
\} (c 2)))
\end{Verbatim}
The first interesting evaluation step is after the underlined \Verb|checkpoint| has been handled.
\begin{Verbatim}[commandchars=\\\{\}]
evaluate (
  reversec (<Smooth(s a -> s)> (reverse (write (deriv (
    let z = prop 4 <r2> in
    let a = checkpoint \{
      let w = checkpoint {t (prop 2 <z>) z} in
      p w (prop 2 <r1>)\} in
    p a (prop 2 <z>)
  )) (<Smooth> (c 1)))));
  reversec (write (deriv (<Smooth(s a b -> s b), RefState> 
                    \emph{\{p (prop 2 <z>) (prop 2 <r1>)\}}!))
                  (read (deriv (prop 4 <r2>))));
  read (deriv (prop 2 <z>)))
\end{Verbatim}
Note how on the second line the \UseVerb{reverse} handler has been made the innermost delimiter of the remainder of the initial program, via the catch-all case of \UseVerb{reversec}.
Additionally, note how the checkpointed code (underlined) is stored as a thunk to be run after the initial program in the second use of \UseVerb{reversec}.
After the initial program has been evaluated away, we obtain the following.
\begin{Verbatim}[commandchars=\\\{\}]
evaluate (
  reversec (<Smooth(s a -> s)> (
    reverse (write <r4> 1);
    write <r3> (p (read <r3>) 
      (t (der2L plusE 10 2) (read (deriv (prop 12 <r4>)))));
    write <z> (p (read <z>) 
      (t (der2R plusE 10 2) (read (deriv (prop 12 <r4>))))));
  reversec (write (deriv (<Smooth(s a b -> s b), RefState> 
      \{let w = \emph{checkpoint {t (prop 2 <z>) (prop 4 <r2>)}} in
       p w (prop 2 <r1>)\}!)) 
    (read (deriv (prop 10 <r3>))));
  reversec (write (deriv (<Smooth(s a b -> s b), RefState> 
      \{p (prop 2 <z>) (prop 2 <r1>)\}!))
    (read (deriv (prop 4 <r2>))));
  read (deriv (prop 2 <z>)))
\end{Verbatim}
The remaining \Verb|checkpoint| command illustrates the recursive nature of \UseVerb{reversec}.
It shows how even nested checkpointing in checkpointed code can be properly evaluated by delaying the program transformation happening via evaluation.

\section{Conclusion}

We have seen the implementation and evaluation of \ac{ad} in Frank via Frank's operation semantics and four handlers: \UseVerb{evaluate}, \UseVerb{diff}, \UseVerb{reverse}, and \UseVerb{reversec}.
While \UseVerb{evaluate} and \UseVerb{diff} do effectively no program transformations, \UseVerb{reverse} and \UseVerb{reversec} build up ancillary programs via delimited continuations.
The effects and handler style of Frank allowed us to compose and nest our defined handlers, which is especially apparent in the modular definition of \UseVerb{reversec} which delegates all \UseVerb{Smooth} commands to \UseVerb{reverse}.
It may also be possible to integrate multi-stage programming by using the system of \citeauthor{wang_demystifying_2019}.
In conclusion, we have illustrated in Frank that effects and handlers are a good match for \ac{ad}, and that effects and handlers can be seen as a form of program manipulation.

\begin{acks}
I would like to thank Sam Lindley for answering my many questions about Frank, my supervisor Chris Heunen for his support, and Ohad Kammar for conversations about this work and encouragement to improve it.
I would also like to thank the reviewers for their valuable feedback.
\end{acks}

\normalem
\bibliography{PEPM'21}

\appendix
\section*{Appendix}
The following is the definition of \Verb|evaluatet| use in \UseVerb{reversec}.
Note the similarities with \UseVerb{evaluate}. 
\begin{Verbatim}
evaluatet : Ref X
         -> <Checkpoint X, Smooth (Prop X)> Y
         -> [Smooth X] Y
evaluatet _ x = x
evaluatet s <checkpoint p -> k> =
  let res = evaluatet s (<Smooth(s a b -> s b)> p!) in
  evaluatet s (k (prop (fwd res) s))
evaluatet s <ap0 n -> k> =
  evaluatet s (k (prop (<Smooth> (op0 n)) s))
evaluatet s <ap1 u (prop x dx) -> k> =
  evaluatet s (k (prop (<Smooth> (op1 u x)) s))
evaluatet s <ap2 b (prop x dx) (prop y dy) -> k> =
  evaluatet s (k (prop (<Smooth> (op2 b x y)) s))
\end{Verbatim}
\end{document}